\documentstyle[preprint,aps]{revtex}

 \newcommand \be {\begin{equation}}
\newcommand \bea {\begin{eqnarray} \nonumber }
\newcommand \ee {\end{equation}}
\newcommand \eea {\end{eqnarray}}
 
 \newcommand \bi {\bibitem}
\newcommand \s {\sigma}

\newcommand \la {\lambda}
\newcommand \La {\Lambda}

\newcommand \Tr {\mbox{Tr}}
\begin{document}
\draft      
\title{Quantum critical effects in mean-field glassy systems}
\author{Felix Ritort}
\address{Institute of Theoretical Physics\\ 
University of Amsterdam\\ Valckenierstraat 65\\ 
1018 XE Amsterdam (The Netherlands).\\ 
E-Mail: ritort@phys.uva.nl}

\date{\today}
\maketitle

\begin{abstract}
We consider the effects of quantum fluctuations in mean-field quantum
spin-glass models with pairwise interactions. We examine the nature of
the quantum glass transition at zero temperature in a transverse
field. In models (such as the random orthogonal model) where the
classical phase transition is discontinuous an analysis using the
static approximation reveals that the transition becomes continuous at
zero temperature. 
\end{abstract} 

\vfill

\vfill

\narrowtext
Spin glasses are models which deserve considerable interest
\cite{BOOKS}. In these systems the presence of randomness and
frustration can yield very rich behaviour. In particular, there is
much current interest in the behaviour of glassy systems in the
presence of quantum fluctuations where the nature of the
zero-temperature phase transition is driven by the competition between
randomness and quantum effects rather than thermal
fluctuations\cite{YoRi}. This makes the order-disorder transition in
quantum glasses belong to a new universality class.

Much work has been devoted to the study of mean-field quantum spin
glass models. In particular, attention has been paid to models with a
continuous transition in the Edwards-Anderson order parameter. The
simplest example in these class of models is the
Sherrington-Kirkpatrick ($SK$) model \cite{SK} in a transverse
field. In this system the critical temperature is depressed when the
transverse field is switched on and vanishes for a critical value of
the field \cite{BM,Usa,IsYa,MiHu}.  Analytical work in the quantum
$SK$ model reveals that replica symmetry is broken in the quantum
glass phase at zero temperature \cite{GoLa}. This is an indication
that quantum fluctuations do not destroy one of the most interesting
features in glassy systems, that is the coexistence of a large number
of phases or states.

There has been also much recent interest in the study of classical
spin-glass models with a discontinuous transition in the order
parameter. These models are characterised by the existence of a
dynamical singularity at a temperature above the static transition
\cite{KiTh}. Concerning the statical and dynamical behaviour these
models are very good candidates for describing real glasses
\cite{Angell}. On the one hand, the statics gives a natural
explanation for the existence of a thermodynamic ideal glass
transition driven by an entropy collapse. On the other hand, the
dynamics of these mean-field models are described by the mode coupling
equations introduced to describe relaxational phenomena in glasses
\cite{KiTh,Angell,FrHe}.  In mean-field models metastable states have
an infinite lifetime, hence dynamics is frozen at the dynamical
singularity well above the static transition temperature. Below the
dynamical transition temperature the system gets trapped in states
with have larger energy than the equilibrium one \cite{CuKu}. All
these features are absent in models with a continuous transition.

The purpose of this letter is the study of models with a discontinuous
transition in the presence of quantum fluctuations at zero
temperature. The motivation is twofold. Concerning the statics we note
that the transition cannot be driven in the quantum case by an entropy
collapse. The reason is that the entropy vanishes everywhere at zero
temperature. Concerning the dynamics we can also expect a quite
different behaviour from the classical case. In macroscopic quantum
systems at $T=0$ the dynamics is governed by the Schr\"{o}edinger
equation and there is no room for any kind of thermal activated
processes.  It could well be that trapping dynamics in the metastable
glassy phase are considerably modified or even suppressed in the
presence of quantum fluctuation effects.

The main conclusion of this work is that the glassy scenario presented
before is indeed suppressed by quantum fluctuations in a certain class
of spin-glass models. We will provide a general proof for this
statement within the static approximation \cite{BM}. In models where
the classical transition is continuous it remains continuous at zero
temperature.

The family of models we are interested in are quantum Ising spin
glasses with pairwise interactions in the presence of transverse
field. These are described by the Hamiltonian,

\be
{\cal H}=-\sum_{i<j}J_{ij}\s_i^z\s_j^z-\Gamma\sum_i\s_i^x
\label{eq1}
\ee

where $\s_i^z,\s_i^x$ are the Pauli spin matrices and $\Gamma$ is the
the transverse field. The indices $i,j$ run from 1 to $N$ where $N$ is
the number of sites. The $J_{ij}$ are the couplings taken from an
ensemble of random symmetric matrices. In the case that the $J_{ij}$
are independent Gaussian variables this Hamiltonian reduces to the
quantum $SK$ model \cite{SK} in a transverse field. If the $J_{ij}$
are orthogonal matrices then eq.(\ref{eq1}) reduces to the random
orthogonal model ($ROM$) \cite{ROM} in a transverse field. At zero
transverse field the models become classical and display quite
different behaviour. The $SK$ model has a continuous finite
temperature transition without jump in the Edwards-Anderson order
parameter \cite{BOOKS} while the $ROM$ presents a strong discontinuous
transition where the Edwards-Anderson order parameter jumps to a value
close to 1 at the transition temperature \cite{ROM}.

In order to solve model (\ref{eq1}) we apply the Trotter-Suzuki
decomposition \cite{TrSu} and rewrite the Hamiltonian in terms of
classical spins with an extra imaginary time dimension,

\be
{\cal H}_{eff}=A\sum_{i<j}\,J_{ij}\sum_t\s_i^t\s_j^t+B\sum_{it}
\s_i^t\s_i^{t+1} + C 
\label{eq2}
\ee

where the time index $t$ runs from 1 to $M$ and the spins $\s_i^t$ take
the values $\pm 1$. The constants $A$, $B$ and $C$ are given by
$A=\frac{\beta}{M}; B=\frac{1}{2}\log(coth(\frac{\beta\Gamma}{M}));
C=\frac{MN}{2}\log(\frac{1}{2} sinh(\frac{2\beta\Gamma}{M}))$. Now we
apply the replica trick and compute the average over the disorder of the
replicated partition function, \be \overline{Z_J^n}=\int [dJ]
\sum_{\lbrace\s_i^t\rbrace} \, exp(\sum_{a=1}^n\,{\cal H}_{eff}^a)
\label{eq3}
\ee where $\int [dJ]$ means integration over the random ensemble of
matrices. This integral can be done using known methods in matrix
theory \cite{ItZu,ROM}. The final result of eq.(\ref{eq3}) can be
written in terms of a generating function $G(x)$ which depends on the
particular ensemble of $J_{ij}$ couplings via its spectrum of
eigenvalues. For the two examples we will consider in this paper we
have $G_{SK}(x)=\frac{x^2}{2}$ ($SK$ model) and
$G_{ROM}(x)=\frac{1}{2}
log(\frac{\sqrt{1+4x^2}-1}{2x^2})+\frac{1}{2}\sqrt{1+4x^2}-
\frac{1}{2}$ ($ROM$ model). From (\ref{eq3}) we get,

\be 
\overline{Z_J^n}=\int dQ\,d\La exp(-N F(Q,\La))
\label{eq4}
\ee
where 
\be
F(Q,\La)=-\frac{nC}{N}+\frac{1}{M^2}\Tr(Q\La)-\frac{1}{2}\,\Tr G(AQ)-\log(H(\La))
\label{eq5}
\ee
with $Q_{ab}^{tt'},\La_{ab}^{tt'}$ being the order parameter 
and the trace $\Tr$ is done over the replica and time indices. The term
$H(\La)$ is given by,
\be
H(\La)=\sum_{\s}\,exp(\sum_{ab}\frac{1}{M^2}\sum_{t t'}\La_{ab}^{t
t'}\s_a^t\s_b^{t'}\,+\,B\sum_{at}\s_a^t\s_a^{t+1}) 
\label{eq6}
\ee and the free energy is obtained by making the analytic continuation
$\beta f=\lim_{n\to 0} \frac{F(Q^*\La^*)}{n}$ where $Q^*,\La^*$ are
solutions of the saddle point equations,
$\La_{ab}^{tt'}=\frac{AM^2}{2}\Bigl (G'(AQ)\Bigr )^{tt'}_{ab}$ and
$Q_{ab}^{tt'}=\langle \s_a^t\s_b^{t'}\rangle$. The average
$\langle(\cdot)\rangle$ is done over the effective Hamiltonian in
(\ref{eq6}). We assume that the order parameters
$Q_{ab}^{tt'},\La_{ab}^{tt'}$ are independent of the time indices when
$a\ne b$ but they are only translational time invariant if $a=b$. To
study the high-temperature phase and the phase boundary of the model
we consider a general one step replica symmetry breaking solution. We
divide the $n$ replicas into $n/m$ boxes $K$ of size $m$ such that $m$
divides $n$. The saddle point solution when $a\ne b$ takes the form
$Q_{ab}^{tt'}=q; \La_{ab}^{tt'}=\la$ if $a,b\in K$ and
$Q_{ab}^{tt'}=\La_{ab}^{tt'}=0$ otherwise. For $a=b$ we take
$Q_{aa}^{tt'}=R_{|t-t'|},\La_{aa}^{tt'}=\La_{|t-t'|}$. Going to the
frequency space we define
$\hat{R}_p=M^{-1}\sum_{t=0}^{M-1}e^{i\omega_p t} R_t$,
$\hat{\La}_p=M^{-1}\sum_{t=0}^{M-1}e^{i\omega_p t} \La_t$ and
$\hat{\s}_p=M^{-1}\sum_{t=0}^{M-1}e^{i\omega_p t} \s_{t+1}$ where
$\omega_p=\frac{2\pi p}{M}$.  The free energy (\ref{eq5}) is given by,

\begin{eqnarray}
\beta f=-\frac{C}{N}+\sum_{p=0}^{M-1} \hat{R}_p\hat{\La}_p^*\,+\,(m-1)q\la
-\frac{1}{2}\sum_{p=0}^{M-1}G\Bigl (\beta\hat{R}_p \Bigr ) 
-\frac{1}{2m}G\Bigl (\beta(\hat{R}_0+(m-1)q)\Bigr )\nonumber\\
-\frac{m-1}{2m} G\Bigl(\beta(\hat{R}_0-q)\Bigr )\,+
\frac{1}{2} G\Bigl (\beta\hat{R}_0 \Bigr )
-\frac{1}{m} log \int^{\infty}_{-\infty} dp(x) \Xi^m(x)
\label{eq8}
\end{eqnarray}
where $dp(x)=(2\pi)^{-\frac{1}{2}} e^{-\frac{x^2}{2}}dx$ is the
Gaussian measure and $\Xi(x)=\sum_{\hat{\s}_p} exp(\Theta(\hat{\s}_p,x))$
with
\be
\Theta(\hat{\s}_p,x)=\sum_{p=0}^{M-1}\Bigl(\hat{\La}_p+MBe^{-i\omega_p}\Bigr)
|\hat{\s}_p|^2 \,+\,(2\la)^{\frac{1}{2}}x\hat{\s}_0-\la\hat{\s}_0^2~~~.
\label{eq9}
\ee

In order to investigate the glassy scenario we compute the static and
dynamical transition temperatures. Following \cite{dynam} we expand the
free energy (\ref{eq8}) around $m=1$, $f=f_0+(m-1)f_1\,+O((m-1)^2)$ and
determine the paramagnetic free energy $f_0$ and the correction
$f_1$. We get,

\begin{eqnarray}
\beta f_0=-\frac{C}{N}\,+\,\sum_p\hat{R}_p\hat{\La}_p^*\,-\,
\frac{1}{2}\sum_{p}G\Bigl (\beta\hat{R}_p \Bigr
)-I(\hat{\La})\label{eq10}\\ \beta f_1=q\la -\frac{\beta q}{2}G'\Bigl
(\beta\hat{R}_0\Bigr)\, +\,\frac{1}{2}G\Bigl (\beta \hat{R}_0
\Bigr)-\,\frac{1}{2}G\Bigl (\beta(\hat{R}_0-q\Bigr)\,\nonumber\\ +\,
I(\hat{\La})-\exp\Bigl(-I(\hat{\La})\Bigr )\int dp(x) \Xi(x) \log
(\Xi(x))
\label{eq11}
\end{eqnarray}

where $I(\hat{\La})=\log(\Xi(\la=0))$.  Note that $f_0$ does not depend
on $q$ and $\la$ as expected for the paramagnetic part of the free
energy. The static and dynamical transition are
obtained by solving the saddle point equations $\frac{\partial
f_1}{\partial q}=\frac{\partial f_1}{\partial \la}=0$. The static
transition appears when the free energy $f$
coincides with the paramagnetic free energy $f_0$, i.e. $f_1=0$.  The
dynamical transition is given by the presence of a soft mode above the
static transition and is obtained by solving the equation
$\Bigl (\frac{\partial^2 f_1}{\partial q^2}\Bigr )
\Bigl (\frac{\partial^2 f_1}{\partial \la^2}\Bigr )-
\Bigl (\frac{\partial^2 f_1}{\partial q\partial\la}\Bigr )^2=0$.

The solution to these equations yields the critical temperature and
the value of the jump of the order parameter $q$ at the transition. In
case the dynamical and the static transition coincide it can be shown
that $q=\la=0$ and the transition is continuous in the order
parameter. These three equations are complemented by the saddle point
equations for the parameters $\hat{R}_p,\hat{\La}_p$
i.e. $\frac{\partial f_0}{\partial \hat{R}_p}=\frac{\partial
f_0}{\partial\hat{\La}_p}=0$.

We now derive a simple expression for the dynamical transition
temperature. The equation for the soft mode can be worked out and one
finds,

\be
\beta^2 e^{-I(\La)}G''\Bigl(\beta(\hat{R}_0-q)\Bigr)
 \int_{-\infty}^{\infty} dp(x) \Bigl(\langle \hat{s}_0^2\rangle-
\langle \hat{s}_0\rangle^2\Bigr )^2 \Xi(x)=1
\label{eq13}
\ee where the average $\langle(\cdot))\rangle$ is taken over the
effective Hamiltonian eq.(\ref{eq9}).  For a continuous transition
$(q=\la=0)$ eq.(\ref{eq13}) can be written in the simple form \be \chi_0^2
G''(\chi_0)=1
\label{eq13b}
\ee where $\chi_0=\beta (\langle \hat{s}_0^2\rangle- \langle
\hat{s}_0\rangle^2)=\beta\hat{R}_0$ is the longitudinal magnetic
susceptibility.  This equation can be solved (for a given $G(x)$) and
yields the critical $\chi_0$. In particular, for the $SK$ model
$G_{SK}''(x)=1$ which yields the result $\chi_0=1$ in agreement with
known results \cite{BM}. In the $ROM$ the only solution to that
equation is $\chi_0=\infty$. For a continuous transition this implies
a divergent susceptibility at the critical field. Using a perturbative
expansion in powers of $1/\Gamma$ it is possible to use 
eq.(\ref{eq13b}) to obtain $\Gamma_c$ with reasonable accuracy \cite{IsYa}.

Now we come to the main result of this paper, namely that at zero
temperature the quantum transition becomes continuous. Then, the
static and the dynamical transition temperature coincide. We first
consider the static approximation where $\hat{R}_p=\hat{\La}_p=0$ for
$p>0$. Putting $R= \hat{R}_0, \La=\hat{\La}_0$ we find the following
saddle point equations for $R,\La,q,\la$,

\begin{eqnarray}
\La=\frac{\beta}{2}G'(\beta
R);~~~~~~~~~~R=<<\frac{sinh(\Xi_0(x))}{\Xi_0(x)}>>_0\nonumber\\
\la=\frac{\beta}{2}(G'(\beta R)-G'(\beta (R-q));~~q=<<\frac{\Bigl
(\int_{-\infty}^{\infty}dp(z) sinh(T)
(\frac{b}{T})^2\Bigr)^2}{\int_{-\infty}^{\infty}dp(z)
cosh(T)}>>_0\label{eq14}
\end{eqnarray}
where 
\be
<<(\cdot)>>_0=\frac{\int_{-\infty}^{\infty}dp(x)
(\cdot)}{\int_{-\infty}^{\infty}dp(x) cosh(\Xi_0(x))}
\ee
with $\Xi_0(x)=(2\La x^2+\beta^2\Gamma^2)^{\frac{1}{2}}$,
$T=(b^2+\beta^2\Gamma^2)^{\frac{1}{2}}$ and
$b=(2(\La-\la))^{\frac{1}{2}}z + (2\la)^\frac{1}{2}x$. 

Exact expressions are also obtained for the free energies $f_0,f_1$ and
for equation (\ref{eq13}).  This set of equations
can be always numerically solved but explicit results can be
analytically obtained in the zero temperature limit. Plugging the
solution $\La=u \beta,\la=v\beta$ into (\ref{eq14})
and performing the
integrals with the saddle point method we find after some lengthy
computations that $u,v$ and the critical field $\Gamma_c$ satisfy the
equations
\begin{eqnarray}
u=\frac{1}{2}G'(\frac{1}{\Gamma_c-2u})\label{eq15a}\\
v=\frac{1}{2}\Bigl
(G'(\frac{1}{\Gamma_c-2u})-G'(\frac{1}{\Gamma_c-2(u-v)})\Bigr )
\label{eq15b}
\end{eqnarray}
It is easy to check that equations (\ref{eq15a}),(\ref{eq15b}) only admit the
trivial solution $v=0$. It is also possible to show that in case $v=0$
also $f_1=0$. Because $q$ and $R$ vanish with $T$ and the free
energy of this solution coincides with the paramagnetic free energy $f_0$
we conclude that the transition becomes continuous at zero
temperature. In order to determine the critical field $\Gamma_c$
we solve eq.(\ref{eq13}) in the $\beta\to\infty$ limit which yields
\be
(\Gamma_c-2u)^{-2}G''(\frac{1}{\Gamma_c-2u})=1
\label{eq16}
\ee with $\chi_0=(\Gamma_c-2u)^{-1}$. Equation (\ref{eq16}) together
with (\ref{eq15a}) determine the value of $u$ and $\Gamma_c$. At the
quantum transition point the internal energy is given by $U=-\Gamma_c$
while the entropy is given by $S=\frac{1}{2}
G(\frac{1}{\Gamma_c-2u})-(\frac{u}{\Gamma_c-2u})+\frac{1}{2}log(\frac{\Gamma_c}{\Gamma_c-2u})$. In
case of the $SK$ model we obtain $u=\frac{1}{2},\Gamma_c=2$
reproducing known results \cite{Usa,KiThLi}. In the case of the $ROM$
we obtain $u=\frac{1}{2},\Gamma_c=1$. Note that in both models the
value of the critical field is given by the maximum eigenvalue of the
coupling matrix $J_{ij}$. For the $ROM$, one finds that the entropy at
zero temperature diverges. This is a consequence of the general
failure of the static approximation at low temperatures. In figure 1
we show the phase boundaries for the dynamical and static transitions
in the $ROM$ as a function of the transverse field obtained
numerically solving equations (\ref{eq14}). Both transition
temperatures decrease quadratically as a function of the transverse
field merging into the same point at zero temperature.  In figure 2 we
show the Edwards-Anderson order parameter $q=\overline{<\s^z>^2}$ in
the $ROM$ as a function of $\Gamma$ as we move along the static
($q_S$) and dynamical ($q_D$) phase boundaries.

The static approximation yields inaccurate quantitative results for
the thermodynamic properties at zero temperature. Nevertheless we
expect the order of the transition to be correctly predicted.  To go
beyond the static approximation we should consider all the Fourier
modes $\hat{R}_p,\hat{\La}_p$ in the saddle point equations. This is a
non trivial task which remains open.

We stress that the main results of this work are restricted to pairwise
interaction models. In case of $p$-spin interaction spin-glass models
\cite{Gar} the scenario presented here does not apply anymore. For $p$
larger than $2$ we have obtained different results indicating that
the transition does not become continuous at zero temperature. This
implies that the quantum transition in $p$-spin interaction models with
$p$ larger than $2$ belong to a different universality class. This is in
agreement with known results in case of $p$-spin models in the
limit $p\to\infty$ \cite{Golds} as well as in $1/p$ analytical
expansions \cite{expans}.

Summarising, we have investigated the glassy behaviour in Ising spin
glass models with pairwise interactions in the presence of a
transverse field. In models with a discontinuous finite temperature
transition we have shown, within the static approximation, that the
transition becomes continuous at $T=0$ and there is no room for a
metastable glassy phase. This implies that at $T=0$ all spin-glass
models with pairwise interactions belong to the same universality
class. 

{\bf Acknowledgements.} I thank D. Lancaster, Th. M. Nieuwenhuizen and
F. G. Padilla for discussions and D. Lancaster for a careful
reading of the manuscript. This work has been supported by FOM (The
Netherlands).

\vfill\eject

\vfill
\newpage
{\bf Figure Captions}
\begin{itemize}

\item[Fig.1] Phase boundaries $T_s(\Gamma)$ (lower line) and
$T_D(\Gamma)$ (upper line) in the $ROM$ in the static approximation.  At
zero transverse field $T_s\simeq 0.0646,T_D\simeq 0.1336$.

\item[Fig.2] Edwards-Anderson parameter $q_s$ (upper line) and $q_D$
(lower line) in the $ROM$ on the static and dynamical phase boundaries
boundaries as a function of the transverse field. At zero transverse
field $q_s\simeq 0.99983,q_D\simeq 0.961$.  $q_s$ and $q_D$ vanish
linearly with $T^{\frac{1}{2}}$ at zero temperature.

\end{itemize}

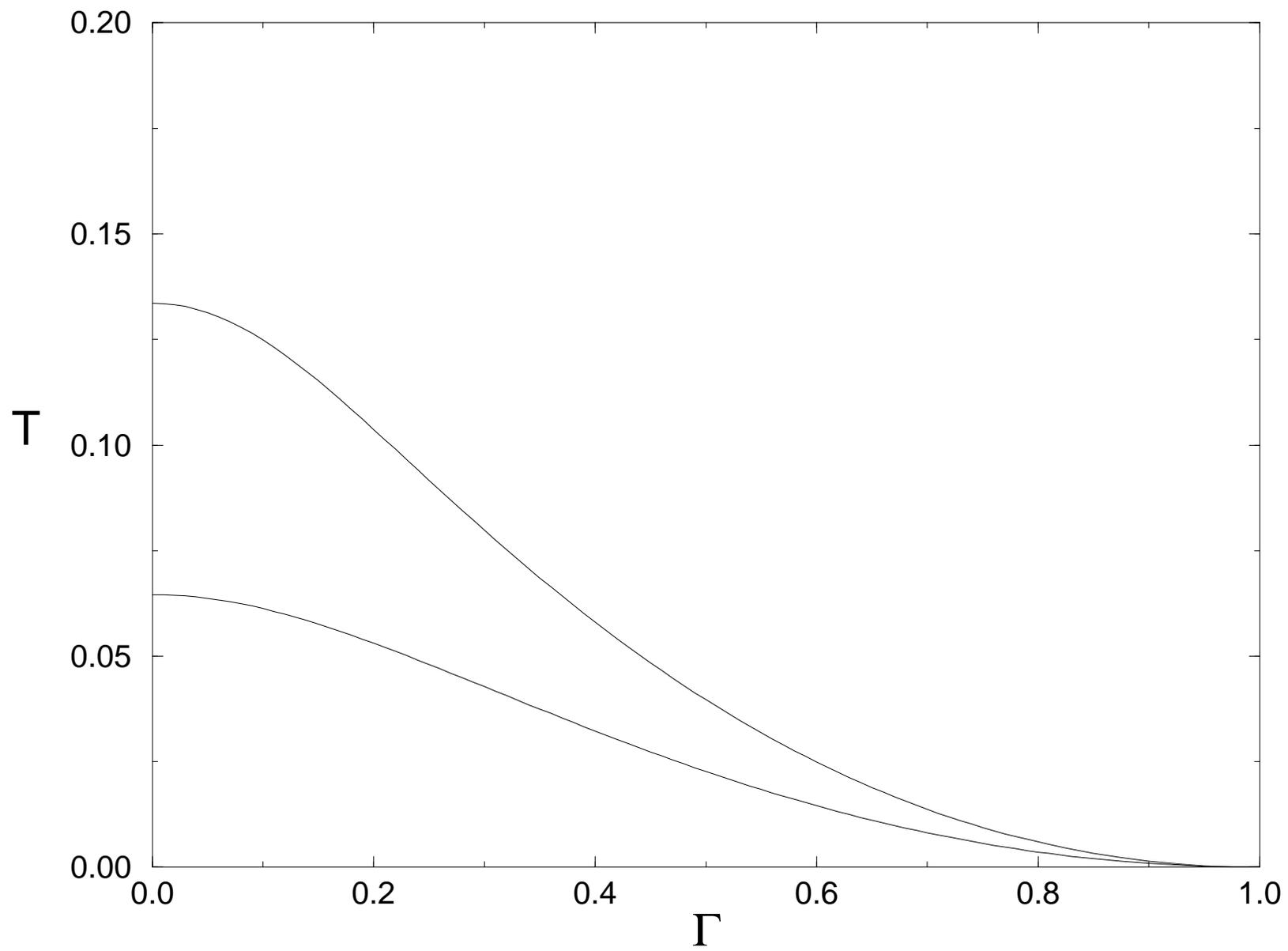
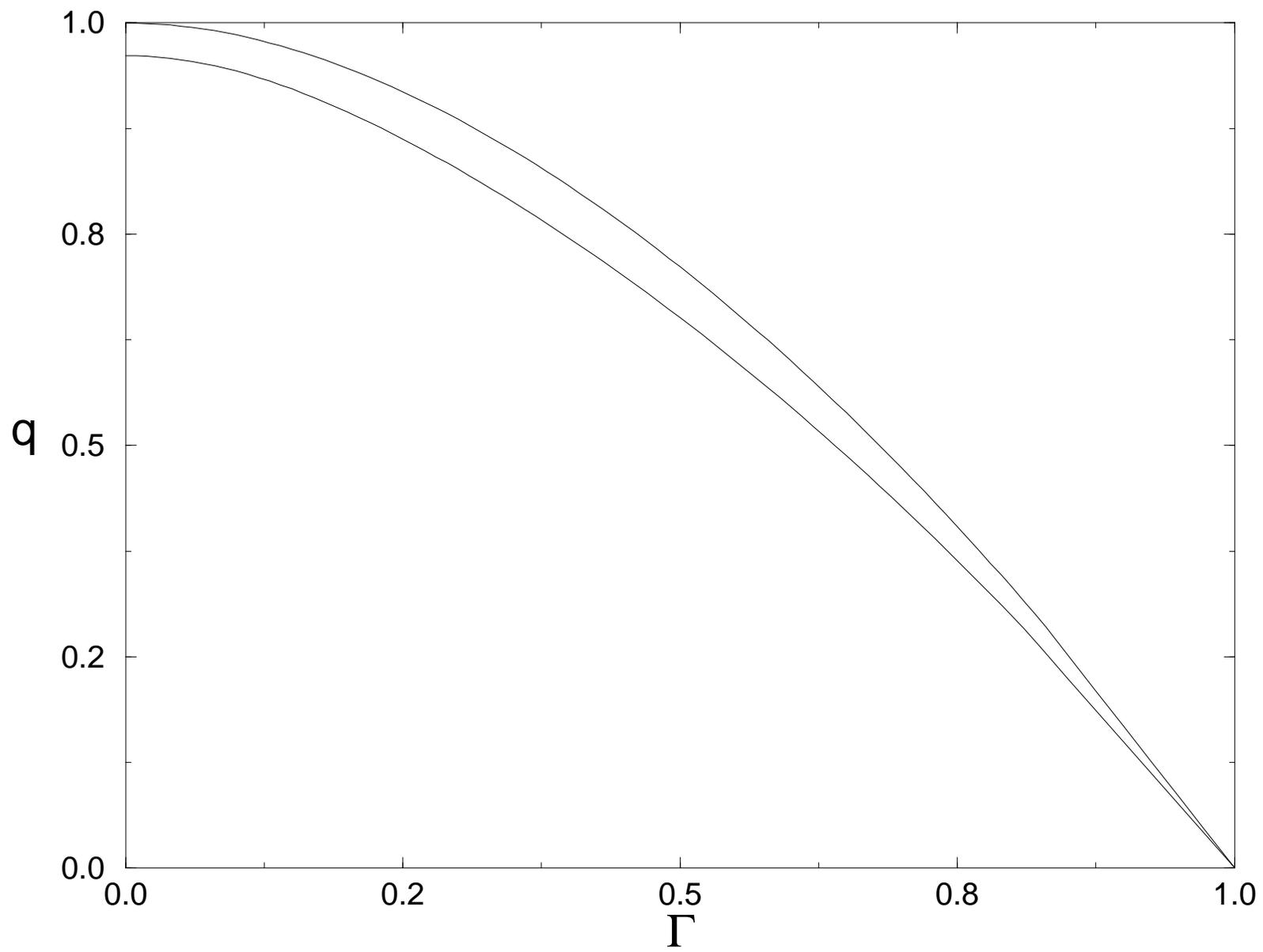
\end{document}